\documentclass[conference, a4paper]{IEEEtran}
\IEEEoverridecommandlockouts

\ifCLASSINFOpdf
  \usepackage[pdftex]{graphicx}
  \DeclareGraphicsExtensions{.pdf,.jpeg,.png}
\else
\fi

\usepackage[colorlinks = true,
            linkcolor = blue,
            urlcolor  = blue,
            citecolor = blue,
            anchorcolor = blue]{hyperref}

\usepackage{cite}

\usepackage{eurosym}
\usepackage{hyperref}
\usepackage{amsmath,amssymb,amsfonts}
\usepackage{algorithmic}
\usepackage{graphicx}
\usepackage{textcomp}
\usepackage[left=1.4cm, right=1.4cm, top=1.7cm]{geometry}
\usepackage{caption}
\usepackage{anyfontsize}
\usepackage{makecell}
\usepackage{arydshln}
\makeatletter
\newcommand*\titleheader[1]{\gdef\@titleheader{#1}}
\AtBeginDocument{%
\let\st@red@title\@title
\def\@title{%
\bgroup\normalfont\normalsize\centering\@titleheader\par\egroup
\vskip0.2em\st@red@title}
}
\makeatother

\makeatletter
\renewcommand{\fnum@figure}{Figure \thefigure}
\makeatother

\title{Determinants of Performance in European ATM\\
\centering{\Large{How to Analyze a Diverse Industry}} 

\vspace{0.5cm}
}

\titleheader{Fifteenth USA/Europe Air Traffic Management Research and Development Seminar (ATM2023)}

\author{\IEEEauthorblockN{Thomas Standfuss, Hartmut Fricke}
\IEEEauthorblockA{Chair of Aviation Technology and Logistics \\
TU Dresden \\
Dresden, Germany \\
thomas.standfuss@tu-dresden.de}
\and
\IEEEauthorblockN{Georg Hirte}
\IEEEauthorblockA{Chair of Economics, esp.\\  Transport Policy and Spatial Economics \\
TU Dresden \\
Dresden, Germany
}

\and

\IEEEauthorblockN{Frank Fichert}
\IEEEauthorblockA{Worms University of Applied Sciences \\
Worms, Germany 
}

}

\IEEEaftertitletext{\vspace{-1\baselineskip}}

\begin{document}

\maketitle

\noindent \begin{abstract}
Air traffic control is considered to be a bottleneck in European air traffic management. As a result, the performance of the air navigation service providers is critically examined and also used for benchmarking. Using quantitative methods, we investigate which endogenous and exogenous factors affect the performance of air traffic control units on different levels. The methodological discussion is complemented by an empirical analysis. Results may be used to derive recommendations for operators, airspace users, and policymakers. We find that efficiency depends significantly on traffic patterns and the decisions of airspace users, but changes in the airspace structure could also make a significant contribution to performance improvements.
\end{abstract}

\vspace{0.3cm}


\section{Introduction}
The efficiency evaluation of air navigation service providers (ANSPs) is still a young discipline in transport and economic sciences. ANSPs represent a natural monopoly and as such are regulated. Therefore, the European Commission set out inter alia cost- and capacity targets. These regulations are valid for well-defined, so-called reference periods (RPs). The recent reference period (RP3) started in 2019 and adopted far-reaching measures to monitor and improve the performance of air navigation services \cite{PRB2018}. 

There are several suggestions on how to improve the efficiency of ANSPs. The most popular example is the 'defragmentation' of European airspace. Studies are predicting potential cost savings due to less fragmentation between \EUR{3.3b} and \EUR{4b} \cite{Adler2022a,EuropeanCommission2023}. However, representatives from politics, industry, and academia have criticized the implementation progress of airspace mergers or cooperations, such as Functional Airspace Blocks (FABs), as far too slow \cite{Commission2012}.  

Even though such criticism can be followed, the reasons for current inefficiencies are clearly complex and multi-layered. Considering that the air transport system consists of different stakeholders with partly contradictory objective functions, an analysis of efficiency drivers should include dependencies of the various endogenous and exogenous factors. Yet, academic research focused on specific/individual mechanisms, rather than providing a holistic approach. In consequence, there is still a lack of a fundamental discussion on potential factors that are influencing performance, the corresponding data and metrics, as well as scientifically sound methods to be applied for a root cause analysis. We aim to close this gap by answering the following questions:
\begin{enumerate}
    \item Which factors influence ANSPs' performance and how can they be measured?
    \item Which methods are suitable to quantify effects on productivity and efficiency?
    \item What actually impacts performance and at what level?
\end{enumerate}

Therefore, the paper is structured as follows: The second section deals with a short overview of fundamental procedures within ANSP operations, which should also be considered in a performance assessment. Based on this, we discuss potential factors in section 3, supplemented by some discussion on how these will probably affect performance. In section 4, we describe the applied method and show the results, which may also be used to derive recommendations to stakeholders. Section 5 summarizes the findings and provides a way forward. The paper refers to \cite{Standfuss2021}.

\section{Background}
\subsection{ANSP Operations in a Nutshell}
\begin{figure}[htb!]
    \centering
    \includegraphics[width=0.48\textwidth]{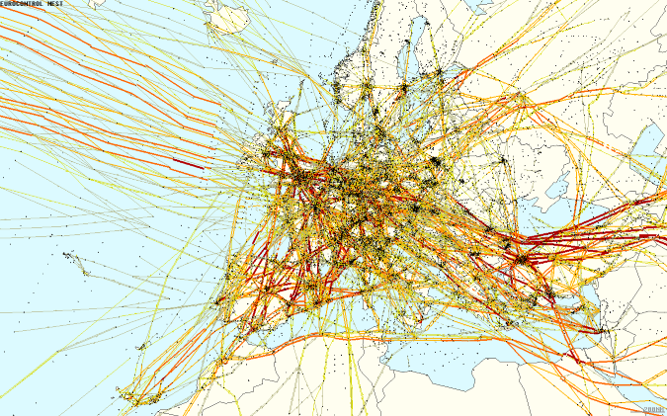}
    \caption{Density Plot of trajectories across Europe, 2019}
    \label{fig:DensPlot}
\end{figure}

European airspace is one of the busiest in the world. Despite a tremendous decrease in traffic due to COVID, demand is meanwhile recovering quickly. However, the spatial traffic load distribution remains odd with a strong shift especially due to the war in Ukraine: The most frequented routes are within the core area of Europe (see \autoref{fig:DensPlot}), where seven large Hubs, such as London Heathrow, Frankfurt, and Paris Charles de Gaulle, are located within a 1,000 km diameter.

In order to cope with the recovery of air traffic demand, ANSPs aim to provide well-adjusted capacity to ensure a safe and efficient traffic flow while aiming to keep operational costs low. The primary goal is to provide a safe and orderly flow of traffic, so as to keep the risk of mid-air collisions as small as possible. Therefore, Air Traffic Control (ATC) separates traffic vertically and horizontally through sequencing techniques. As a consequence of the traffic density shown in \autoref{fig:DensPlot}, ANSPs located in the core of Europe are faced with significant peak demand figures and as such with challenging capacity management.

\begin{figure*}[htb!]
    \centering
    \includegraphics[width=0.66\textwidth]{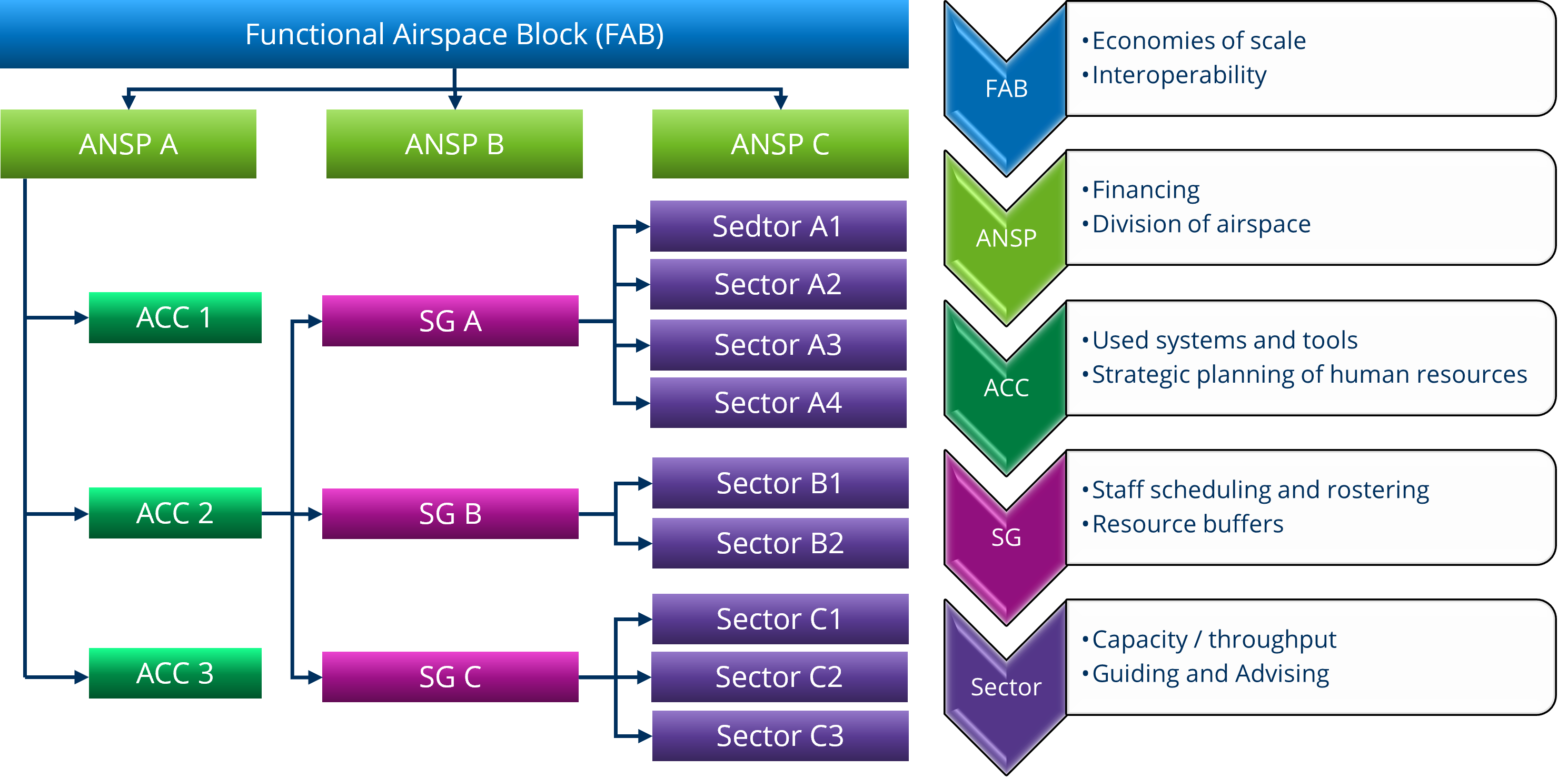}
    \caption{Operational Levels, Objectives and Determinants (Enroute Air Traffic Control)}
    \label{fig:Opslevel}
\end{figure*}

Organizationally, air traffic control is divided into ’terminal’ and ’enroute‘ services. They differ significantly in their operational procedures. Further, in order to manage the challenges of air traffic complexity, especially in congested areas, the enroute part is laterally split into multiple operational levels, these are Area Control Centers (ACC), themselves comprising multiple sector groups (SG), and further sectors, as shown in \autoref{fig:Opslevel}. Each disaggregation level is characterized by specific objectives and is subject to constraints as well as environmental influences. 
 
Depending on the size of the ANSP, enroute services are provided in one or more ACCs to cover a specific part of the airspace. These (geographic) areas are designed following different characteristics, e.g., traffic flows or specific altitudes \cite{FABEC2019}. Usually, ACCs are responsible for both upper and lower airspace. However, there are exceptions: Karlsruhe and Maastricht, for example, only provide services for the upper airspace, which is why they are also referred to as Upper Area Control Centre (UAC). 

The number of acceptable operations within a sector is mainly determined by the Capacity Default Value (CDV). It is either defined by aircraft entries per hour or by maximum occupancy counts\footnote{Number of flights simultaneously operating in a given area}. Further levels of disaggregation are sector groups and sectors. The latter represent the smallest operational unit, where ATCOs are responsible for all aircraft in a specific geographical area. 

It is a common practice to split or collapse (merge) sectors over time to align capacity to current demand following sector opening/closing schemes or configurations, respectively. Since sector size does not correlate linearly to capacity, we found a diminishing marginal capacity increase by sector splitting\cite{Standfuss2018d}. Furthermore, collapsing sectors are operationally limited to SGs, for which controllers hold a license. Those groups are consequently referred to as licensed areas. %As an example and referring to \autoref{fig:Opslevel}, sector C1 could be merged with sector C2, but not with sector B2.

\subsection{Literature review}\label{lit}
The academic literature provides various approaches to assessing the performance of decision-making units (DMU). In economics. a common practice is the so-called "two-stage analysis". In the first stage, the performance of the unit is determined by means of a previously defined metric. The second stage then examines which influencing factors exist and how they affect the performance score in terms of sign (does the factor have a positive or negative effect on the performance), strength (how high is the influence), and statistical significance (is this relationship statistically verifiable or rather random).

The analysis of ANSP performance (first stage) was introduced some 20 years ago by the European Organisation for the Safety of Air Navigation EUROCONTROL, and disseminated in annual reports, e.g., \cite{EUROCONTROL2019g}. Academic studies particularly improved the assessment methodology, applying e.g., data envelopment analysis (DEA) \cite{Bilotkach2015,Standfuss2022}, or stochastic frontier analysis (SFA) \cite{Blondiau2016}. 

Since DEA is a frequently used methodology to determine efficiency (first stage), various publications deal with the appropriate second-stage method. \cite{Simar2007} analyzed and compared different regression techniques, including Ordinary-Least-Squares (OLS), Truncated and Tobit models. They also tested different data transformations, particularly the log-transformation of DEA values. The authors recommend a truncated model for regression based on a DEA. Further, \cite{Banker2008} compared different second stage approaches. The authors showed that deterministic methods are superior to parametric approaches. The most common approach is applying DEA in combination with a Tobit regression, e.g., \cite{Spaho2015}, although \cite{Hoff2007} demonstrated that these models perform similarly to OLS models.

Further academic studies primarily dealt with specific aspects of efficiency influencing factors (second stage) and how to improve efficiency. As an example, \cite{Starita2021,Button2013,Standfuss2019f} analyze potential improvements by achieving economies of scale through airspace mergers. Other studies examined the efficiency gains through the privatization of ANSPs \cite{Buyle2022}, alternative financing concepts \cite{Verbeek2017}, dynamic sectorization, \cite{Gerdes2018}, flight-centric ATC  \cite{Birkmeier2014, Nevir2022}, or alternative market designs\cite{Adler2022}. 

Operational and academic studies usually assess at ANSP level\footnote{Since most data (particularly financial data) is available only on this level}. On disaggregated level, some investigations addressing fundamental aspects of capacity provision in ACCs \cite{FABEC2021} or specific efficiency drivers \cite{FABEC2018b}. Further, \cite{Standfuss2017b} provided a methodical approach how to benchmark performance at disaggregated levels. 

\subsection{Scope and Delimitation of the Study}
The analysis of specific measures, such as merging airspace structures, provides a first valuable insight. However, these studies have focused on the contribution of ANSPs, only. This may be expedient from a regulatory point of view, but neglects influences that are not under the control of these service providers. Obviously, recommendations listed in the literature differ depending on the focused research area, the used data, the developed (economic) models, and the applied methodology\footnote{Same applies to the potential savings through the proposed measures}. In consequence, there is no consistent view on what actually impacts performance. The present paper aims to overcome this gap by holistically considering all factors influencing ANSP performance.

A such holistic investigation shall include all operational levels and multiple key performance areas (KPAs) such as safety, cost, capacity, and environment \cite{EUROCONTROL2021b}. For the sake of robust findings, we limit ourselves as follows:
\begin{itemize}
    \item We address the ANSP level, only.
    \item "Performance" is either expressed by productivity or efficiency scores (see \autoref{sec:appr}). We do not consider e.g., environmental indicators such as HFE \cite{EUROCONTROL2021} or XFB \cite{EUROCONTROL2020q}.
    \item We use performance scores provided by EUROCONTROL \cite{EUROCONTROL2020e} or taken from other publications \cite{Standfuss2021}.
    \item We only use cross-sectional data and models. Panel data and regression were and will be subject to further research.
    \item We consider both endogenous and exogenous effects.
    \item We do not apply data transformation\footnote{Data transformation may help improving model quality. In some cases, coefficients show the relative influence of each factor. Log-transformation is required in case the dependent variable is not distributed normally} (standardization, unitization, etc.) unless stated otherwise.
    \item Only quantitative methods are applied.
\end{itemize}
For those constraints set, we aim to provide a fundamental comparison between different methods as well as an empirical validation. Our investigation is based on 2016 data unless stated differently.

\subsection{Definitions and Approach}\label{sec:appr}
The performance (productivity or efficiency) of an ANSP can be influenced by a variety of factors. These are initially divided into endogenous effects (which can be influenced by the ANSP) and exogenous effects (which in turn cannot be influenced by it). In our study, productivity is defined following EUROCONTROL Performance Review Unit (PRU) scoring. It equals the ratio of Composite Flight hours (Services) and ATCO work hours (resources) \cite{EUROCONTROL2020i, EUROCONTROL2021g}. (Cost) Efficiency either expresses providing services at minimum costs or to produce a fixed output with a minimal deployment of resources (or vice versa). Consequently, the performance of one unit is compared to the best-in-class unit \cite{Coelli2005}. Both metrics have some advantages and disadvantages with regard to completeness, interpretability, and implications on applicable methods \cite{Welc2017,Urban2018}. Thus, we apply both metrics in separate analyses.

\begin{figure*}[htb!]
    \centering
    \includegraphics[width=0.98\textwidth]{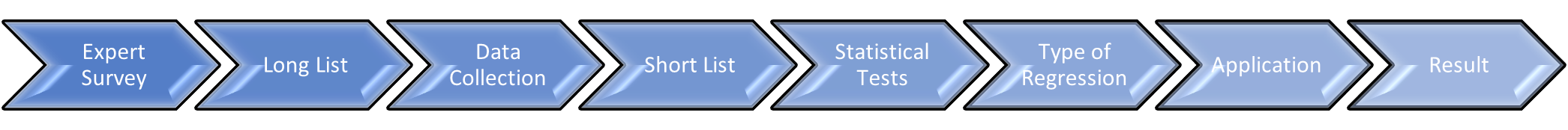}
    \caption{Analysis Steps of a Second-Stage Analysis}
    \label{fig:Steps}
\end{figure*}

\autoref{fig:Steps} shows our procedure for the analysis which consists of eight steps in total. In the first two steps, we identify potential factors impacting performance (long list) based on an expert survey. These factors shall be quantifiable and hold candidate or recognized metrics (step three) to create a short list\footnote{Usually, some factors cannot be included because either data are missing or influences can only be determined qualitatively} (step four). Identification and selection is crucial. Excluding relevant factors will lead to biased results (known as omitted variable bias, OVB). This is in particular true for over- or underestimated variable coefficients. 

Statistical tests (step five) help to identify those variables that might be used in parallel\footnote{In one model} for investigation. These tests might be performed in advance (e.g., correlation analysis, principal component analysis\footnote{Recommended in case of a relatively low number of observations in comparison to the number of potential factors}, etc.) \cite{Gehrke} or might be implemented in the quantitative analysis itself (see e.g., Lasso Regression).

In step six, we apply the proposed quantitative method: regression analysis (see also \autoref{sec:meth}). The type of regression is determined by the specification of the data. Steps seven and eight represent the application of the selected regression technique and the discussion of the results. For sensitivity analysis, we run different models (selection of variables) and/or specifications (e.g., type of regression). We elaborate on the methodological process in \autoref{sec:Regression}.

\section{Potential Efficiency Drivers}
\subsection{Selection and Grouping of Influencing Factors}
As illustrated in \autoref{fig:Steps} as the first step, we identify potential factors which influence the performance of an ANSP. Therefore, we primarily used expert consultations and recent academic publications to create the long list. In addition to the identification of potential factors, the acquisition of the necessary data is an essential part of the investigations (see also \autoref{sec:data}). The short list contains all potential influencing factors that can be quantified using meaningful metrics, and for which data are available. 

To provide a basis for identifying measures for performance enhancement, we categorize these factors into endogenous, partially exogenous\footnote{Meaning that an ANSP has some influence on those factors, but there are exogenous influences as well.}, and exogenous elements. Further, the partially exogenous effects are subdivided with respect to airspace structure and demand characteristics. Although the categorical classification of factors is not always unambiguous (e.g., \textit{JSC} and \textit{STATE} might also be classified as partly-exogenous), the distinction helps to address potential measures to different stakeholders, e.g., airspace users, service providers, and policy decision-makers.

\autoref{tabreg} summarizes the short-listed factors (acronyms) used in the quantitative analysis, which are explained in more detail in the following subsections. The dashed lines represent the distinction between endogenous, partly-exogenous, and exogenous effects. To avoid scaling errors, factors with high absolute expressions of the values (e.g., airspace size) are log-scaled (see index "l" tagged to the acronym). Furthermore, the metric (M) units of the parameters as well as the expected influence (E.I.) on productivity or efficiency are indicated ("+", "-" or "?" for either way). It does not reflect the severity of the impact, but the (expected) sign of the coefficients. This procedure helps both to check the plausibility of the results including any necessary adjustments to the model, and to discuss the results. Since factors may correlate, we will check potential combinations in \autoref{sec:data}. 

\begin{table}[htb!]
    \centering
        \caption{Factors used in the Quantitative Analysis}
    \begin{tabular}{lllc}
\hline
Parameter    & M   & Meaning                                                          & E.I. \\ \hline
TIME (l)     & h     & Working time per ATCO per year                                    & +        \\
NONA         & \%    & Share of Non-ATCOs                        & -        \\
DELATM       & 0/1   & Delegated ATM                                               & ?        \\
MET          & 0/1   & Provision of Meteorological Services                                                       & +        \\
AIRP         & 0/1   & Ownership / Management of Airports                       & +        \\
JSC          & 0/1   & Joint Stock Company                                                & +        \\
STATE        & 0/1   & State-Owned                                                        & -        \\ \hdashline
SIZE (l)     & Km$^2$   & Airspace size                                               & +        \\
OCEAN        & 0/1   & Airspace is (partly) over the ocean                            & ?        \\
COORD        & Nb & Coordination with adjacent units             & -        \\
L\_AIRP      & Nb & \makecell[l]{Number of large airports \\ (\textgreater 200.000 movements p.a.)} & -        \\
NOFAB\ & 0/1   & The ANSP is not organized in a FAB                                      & ?        \\
OVER         & \%    & Share of overflights                                              & +        \\
DOM          & \%    & Share of domestic flights                                            & -        \\
GINI         & \%    & Seasonal volatility of flights                                  & -        \\
DENS      & Score & Adjusted density                               & -        \\
VI           & Score & Vertical interactions                      & -        \\
HI           & Score & Horizontal interactions                    & -        \\
SI           & Score & Speed interactions                & -        \\ \hdashline
COSTS        & \EUR{}   & costs per ATCO hour                                     & +        \\
RES          & Nb & Technology Proxy (Research)                  & +        \\
WEALTH (l)   & \EUR{}     & GDP per capita    & + \\ \hline
    \end{tabular}
    \label{tabreg}
\end{table}

\subsection{Endogenous Factors}
The central aim of the present ANSP performance assessment is to identify potential measures for improvement. Thus endogenous factors are crucial. Here, the human resources, the ATCOs, play a paramount role: We consequently use two variables to describe staff allocation and scheduling. The first variable depicts the average ATCO yearly working time: (\textit{TIME}). As shown in \autoref{tabreg}, we find significant pan-European differences for it. A low value implies more controllers (FTEs) to achieve the same outcome. We consequently assume a positive correlation with performance. \textit{NONA} indicates those employees not performing air traffic control. Those may be seen as support (e.g., flight data analysts), overhead or administration. Since  overhead increases unit costs, we expect the variable to correlate negatively with the productivity and efficiency scores.

\textit{DELATM} and \textit{MET} reflect "make or buy" decisions. An ANSP may outsource ATM services partially (especially ATC), probably because of a given traffic flow pattern. For example, parts of the airspace of the ACC Rome (ENAV) is delegated to the Maltese MATs. This airspace is particularly frequented by flights to and from Malta. If financial compensation is granted for a delegation, the dummy gets the value "1", otherwise "0". The (expected) impact on productivity is ambiguous. A main disadvantage is, that the dummy does not reflect the size of the delegated airspace\footnote{According to \cite{EUROCONTROL2012a}, the size of the airspace associated to an ANSP should be corrected by the size of the allocated airspace. However, it is not clear whether this is done by all ANSPs}, or how many flights are affected. Further, we assume a positive sign for the variable \textit{MET}, since especially large ANSPs benefit from an internal weather bureau. Thus, relevant data on the current weather are available more quickly. In addition, external providers may have limitations with regard to the transmission of data. An even better approximation might be possible using MET costs, but the data is incomplete.

\textit{AIRP} indicates whether the air navigation service provider is also responsible for airport management. This is the case, for example, with Avinor (Norway). The participation in or ownership of airports by the air navigation service provider can facilitate coordination between the two entities.

The performance of an ANSP can also be influenced by the ownership\footnote{The type of ownership is considered to be an endogenous factor, although governmental influences (e.g., regulations) theoretically also may imply exogenous effects.} structure. Economic analyses distinguish between state-owned, privatized, and partially privatized companies. In economics, it is often assumed that the privatization of state-owned companies always leads to higher efficiency \cite{Button2010}. However, \cite{Liebert2011} demonstrated that this is not true for all cases. One example are airport companies, where mixed forms lead to lower efficiency. EUROCONTROL distinguishes between six different types of companies, including "state-owned" (\textit{STATE}) or "Joint Stock Company" (\textit{JSC}) \cite{EUROCONTROL2019g}. Both forms represent a total share of 75\% of ANSPs.  Since we use dummies instead of a "degree" of privatization (due to missing data), we are only able to show whether an organizational form affects performance positively or negatively. According to the literature, a negative influence is expected for \textit{STATE}, while a positive correlation is expected for \textit{JSC}. We argue that it is generally difficult to distinguish between state-owned and privatized ANSPs, since only the British NATS and the Swiss skyguide are considered partially privatized\footnote{Although the private share in skyguide is only very minor}. In addition, there are publications addressing this particular aspect \cite{Buyle2022, Adler2017}.

\subsection{Partly Endogenous Factors}
Some of the factors presented in \autoref{tabreg} could be influenced by the ANSP to a certain level, while still being dependent on other stakeholders or environmental determinants. As an example, we consider airspace as partly endogenous. While ANSPs have some design flexibility in structuring their airspace, the actual size (\textit{SIZE}) is given by the state borders, as long as no FAB is considered. We assume a positive correlation between size and performance since large airspace holds a larger disaggregation potential, which may increase  ATC resilience against traffic fluctuations (e.g., due to incidents or weather). Further, during low traffic peak times (e.g., at night), merging sectors allows a larger area to be covered by a given, limited number of resources (normally two ATCOs). 

The territory of the state may include oceanic airspace (e.g., true for Spain, Portugal, and Ireland), published in official EUROCONTROL reports \cite{EUROCONTROL2019g}. The dummy variable \textit{OCEAN} indicates whether or not the airspace of the corresponding ANSP includes "Oceanic Airspace". It is not clear how oceanic airspace affects performance. On the one hand, large areas (airspaces) must be covered, some of which have little or no traffic. This should have a negative impact on performance. On the other hand, the complexity in these airspaces is low, which has a positive effect on performance.

The location of an ANSP (e.g., within Europe) influences the coordination effort (\textit{COORD}). The more adjacent units, the higher the need for coordination. Further, interoperability between systems is not given necessarily, e.g., in case the neighboring ANSPs are organized in different FABs and/or differ significantly in their determinants and the associated requirements for systems and tools. This was particularly emphasized in \cite{FABEC2019}. Therefore, a negative relationship is expected.

Another potential factor affecting performance is represented by the number of large\footnote{We consider all airports with more than 200,000 aircraft movements per year as "large". This limit corresponds to the top 20 airports in Europe.} airports. This is defined by the variable \textit{L\_AIRP}. Traffic is (spatially) concentrated in areas around the major airports. This increases the complexity, respectively the traffic density and thus the taskload of the controllers. Further, the complexity of the specific runway system may affect performance \cite{Gelhausen2019}. Therefore, a negative correlation is assumed. 

A further aspect of the location is the association to a FAB, which cannot be influenced by an ANSP. However, it would be possible, for example, to adjust the number of ANSPs per FAB. Thus, it is partly exogenous. Since FABs are expected to achieve economies of scale, a positive influence can be assumed. Conversely, this means a disadvantage for ANSPs that are not coordinated in a FAB, expressed in the dummy \textit{NOFAB}. However, \cite{Standfuss2019f} showed, that not each ANSP would benefit by being a member of a FAB. Subsequently, we also tested the association with a specific FAB. It should be noted that \cite{Adler2022a} and \cite{Starita2021} discuss FAB-independent airspace mergers. 

The actual amount and distribution of traffic are largely exogenous. Three main elements are examined in the regression: the type of flight, traffic variability, and complexity. The variable \textit{OVER} represents the share of overflights, and the variable \textit{DOM} the share of domestic flights. Due to collinearity, international flights (take-off or landing in the territory of the ANSP) are not taken into account. We expect that the variable \textit{OVER} has a positive effect on the efficiency, while \textit{DOM} affects performance negatively.

Traffic fluctuations may affect performance since the increased need for resources leads to higher costs. Efficiency losses are expected particularly in the case of a lack of flexibility in staff scheduling. Traffic volatility got into the focus of researchers recently. The definition, calculation, and impact of these fluctuations are addressed in e.g., \cite{Deltuvaite2018, Standfuss2021b}, the latter showed, that the GINI coefficient is valuable to express volatile traffic patterns. Thus, we included the seasonal fluctuations by the variable \textit{GINI}.

Complexity is expected to influence efficiency significantly. Due to data availability, we use the metric provided by the PRU \cite{EUROCONTROL2020c}, considering the individual components \textit{DENS}, \textit{HI}, \textit{VI}, and \textit{SI} for the regression analysis. In general, high complexity implies a higher workload for the controller. This has a negative impact on the capacity of the sector and thus on productivity or efficiency. Therefore, a negative correlation is expected for all four components.

\subsection{Exogenous Factors}
Exogenous effects, such as socio-economic factors like the size of the economy or the prosperity of a country, cannot be influenced by the ANSPs. Other potential factors are weather, geographical location, or other environmental factors. The underlying assumption is that the prosperity (\textit{WEALTH}) of a country is also reflected in the efficiency of the associated ANSP\footnote{e.g. in terms of investment potential}.

The variable \textit{COSTS} describes the labor costs per ATCO hour, which are highly dependent on the wage level of the country. Assuming that high costs lead to more efficient deployment of resources, this variable is expected to have a positive impact on efficiency\footnote{This is not the case for cost-efficiency}. In a previous study, costs were additionally divided by individual purchasing power parities \cite{EuropeanCommission2018}. We did not adjust the cost since wage effects should be considered. Moreover, the use of the PPP index is not uncontroversial \cite{Ong2003}. Efficiency can also depend on the technology level of a country. \textit{RES} reflects patents per 100,000 inhabitants. A positive influence on efficiency is expected.

\section{What Actually Influences ANSP Performance}\label{sec:Regression}
\subsection{Data}\label{sec:data}
As discussed in \autoref{sec:appr}, potential factors need to be backed up by suitable metrics. This might be challenging in the context of air traffic control since data is in some cases sensitive and/or not published. This applies i.a. to military activities. Other data are not fully available, either in terms of the studied period or the ANSPs considered. 

In addition to transport-specific data, socio-economic data such as wealth or technology may also play a role, as shown in \autoref{tabreg}. Thus we particularly used data provided by \cite{EUROCONTROL2020d,EUROCONTROL2019o,Worldbank2019,EUROSTAT2019}. Since the list of all potential metrics included 51 items, but there is a limited number of observations, we considered variable reduction methods. We applied principal component analysis (PCA) and ran a regression analysis based on the components.  However, the individual factors could not be meaningfully combined into components. Subsequently, the model quality of the regression as well as the statistical significance of the components were insufficient. Despite various adjustments and testing of alternatives, including increasing the number of main components and substituting parameters, no improvement in the result could be achieved. Therefore, regression analysis was performed using the factors themselves. 

\autoref{stats2016} summarizes some of the explanatory variables we used in regressions (excluding dummy variables) for the year 2016. The large scatter of values is due to the heterogeneity of the operational, geographic, and socioeconomic environmental factors of the individual ANSPs. For airspace, the maximum value of 2,190,000 km$^2$ (Spain) can be interpreted as an extreme value. Extreme values are also visible for the factors representing the share of domestic flights and wealth. In addition to these factors, we used several dummy variables. This applies e.g., to the organizational form and the association with FABs. For FABs, both individual affiliation (FABEC, FABCE, etc.) and general affiliation (FAB yes/no) were checked.

\begin{table}[htb!]
\centering
\caption{Descriptive Statistics 2016}
\begin{tabular}{lrrrrr}
 \hline 
Indicator & Min & Median & Max \\
 \hline 
Working time per ATCO (h/a) & 934    & 1,434   &1,990     \\ 
Share of Non-ATCO FTEs      & 38 \%  & 64 \%   &  87 \%    \\ \hdashline
Airspace Size (km$^2$)         & 20,400 & 346,350  & 2,190,000  \\
Coordination                & 2      & 6         & 11           \\
Number of Large Airports               & 0      & 1           & 3         \\
Share Overflights            & 10 \%  & 59 \%   & 100 \%    \\
Share Domestic Flights         & 0 \%   & 7 \%      & 50 \%      \\
Traffic variability (GINI)                       & 2.5 \% & 10.4 \% &  25.5 \%   \\
Traffic Density              & 0.69   & 6.26   & 11.47       \\
Vertical Complexity            & 0.04   & 0.15    & 0.38      \\
Horizontal Complexity          & 0.27   & 0.46    & 0.63       \\
Speed Complexity       & 0.04   & 0.19     & 0.45       \\ \hdashline
Costs per ATCO-hr (\EUR{}) & 11     & 91      & 225        \\
Technology           & 0.26   & 13.33     & 95.76     \\  
Wealth (\EUR{})                   & 2,074  & 29,864   & 181,647   \\
\hline
\end{tabular}

\label{stats2016}
\end{table}

\begin{figure*}[htb!]
    \centering
    \includegraphics[width=0.70\textwidth]{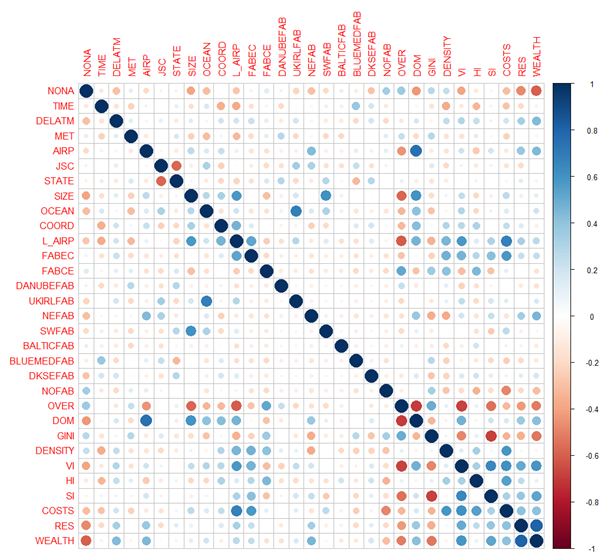}
    \caption{Correlation Plot of all used independent variables)}
    \label{fig:Correl}
\end{figure*}

The shortlisted factors may correlate with each other. It should be avoided to include correlated factors in one model. Instead, one should test these factors sequentially and compare significance and model quality. \autoref{fig:Correl} shows a correlation plot of the shortlisted factors. As one example, there is a high correlation between the share of domestic flights and whether an ANSP also operates airports. However, it might make sense to consider both: Although these data are correlated, it is most likely a spurious correlation. In contrast, the data of \textit{WEALTH} and the technology proxy are also highly correlated. In this case, a connection is plausible, and subsequently, both factors should not be used in parallel in one model.

\subsection{Methodology}\label{sec:meth}

The necessity of considering multiple influencing factors leads to the application of methods like regression analysis. %It allows the quantification of the influence of one or more independent variables (factors) on the chosen dependent variables. The regression calculates how to weight the independent variable(s) (“coefficients”) to estimate the dependent variable(s) (in our case Productivity or Efficiency) as precisely as possible. This accuracy of the regression model is evaluated by model quality criteria. Good model quality is e.g., expressed by a high coefficient of determination (R$^2$): The closer the indicator is to 100\%, the more variance is resolved by the considered factors, respectively by the implied model. 
%The type of regression is dependent on the characteristics of the dependent variable and the data (cross-sectional versus panel data). Cross-sectional data means that data is available for one time period and all firms. Subsequently, no time effects have to be considered. Panel data requires figures for multiple years. Thus, more observations are available, enabling the consideration of a higher number of factors. 
Depending on data characteristics, different regression types might be applied. If panel data is available, panel regression models like Pooled, Fixed- or Random-Effect Models are able to consider (unobserved) time effects. The underlying analytical procedures are well documented e.g., in \cite{Wooldridge2002}, and \cite{Hsiao2014}. For cross-sectional data, Ordinary-Least-Squares regression is the most common method. However, if the dependent variable is restricted, OLS is model misspecification, and Tobit- or Truncated models are to be preferred \cite{Welc2017}. 

There are mainly two ways to apply regression. The most common method is to maximize model quality (e.g. adjusted R$^2$) by variable reduction. That means, that statistically insignificant variables are excluded successively from the regression model. We used variable reduction with a p-value threshold of 0.33. The threshold is arbitrary, but previous studies showed, that extending the threshold inhere some benefits, especially when there are variables that have a p-value of slightly above 0.1. 

Another option is to successively include endogenous, partially exogenous, and exogenous factors. For this purpose, we use five iterations: In the first iteration, only endogenous factors are considered. The second step includes the yearly dummies. Iterations 3 and 4 add airspace data as well as demand characteristics (and thus partially exogenous factors). In the last iteration, all factors are considered. After each intermediate result, a VIF-test is performed to avoid multicollinearity. 

\subsection{Results}

The official EUROCONTROL reports particularly examine ATCO productivity (Composite Flight Hours per ATCO hour). Thus, our first objective is to determine the influencing variables on this metric. For this purpose, the regression model is adjusted by means of variable reduction so that the dependent variable can be predicted as precisely as possible. Since the dependent variable is not upper-bounded, OLS regression is chosen. The results are presented in \autoref{regprod}. 

\begin{table}[htb!]
 \caption{Regression Results PRU Productivity Score}
    \centering
    \begin{tabular}{lcc}
    \hline
    & Coefficient \\ \hline 
INT      & \begin{tabular}[c]{@{}c@{}}-2,69    \\ (1,684)\end{tabular}     &  \\ \hdashline
TIME (l) & \begin{tabular}[c]{@{}c@{}}0,213    \\ (0,178)\end{tabular}     &  \\ 
NONA     & \begin{tabular}[c]{@{}c@{}}-0,454    \\ (0,328)\end{tabular}    &  \\ 
DELATM   & \begin{tabular}[c]{@{}c@{}}-0,138    \\ (0,071*\end{tabular}    &  \\ 
AIRP     & \begin{tabular}[c]{@{}c@{}}0,576    \\ (0,142)***\end{tabular}  &  \\ 
JSC      & \begin{tabular}[c]{@{}c@{}}0,129    \\ (0,073)*\end{tabular}    &  \\ 
STATE    & \begin{tabular}[c]{@{}c@{}}-0,065    \\ (0,065)\end{tabular}    &  \\ \hdashline
SIZE (l) & \begin{tabular}[c]{@{}c@{}}0,125    \\ (0,038)***\end{tabular}  &  \\ 
OCEAN    & \begin{tabular}[c]{@{}c@{}}0,211    \\ (0,095)**\end{tabular}   &  \\ 
COORD    & \begin{tabular}[c]{@{}c@{}}0,019    \\ (0,018)\end{tabular}     &  \\ 
OVER     & \begin{tabular}[c]{@{}c@{}}0,587    \\ (0,15)***\end{tabular}   &  \\ 
DOM      & \begin{tabular}[c]{@{}c@{}}-1,8    \\ (0,568)***\end{tabular}   &  \\ 
GINI     & \begin{tabular}[c]{@{}c@{}}-3,069    \\ (0,597)***\end{tabular} &  \\ 
DENS  & \begin{tabular}[c]{@{}c@{}}0,026    \\ (0,013)*\end{tabular}    &  \\ 
HI       & \begin{tabular}[c]{@{}c@{}}0,619    \\ (0,361)\end{tabular}     &  \\ \hdashline
COSTS    & \begin{tabular}[c]{@{}c@{}}0,002    \\ (0,001)**\end{tabular}   &  \\ 
RES      & \begin{tabular}[c]{@{}c@{}}-0,003    \\ (0,002)\end{tabular}    &  \\ \hline
Korr. R-square & 0,86                                                              &  \\
Akaike   & -36,38                                                            &  \\
N        & 38                                                                & \\
\hline
    \end{tabular}
      \label{regprod}
\vspace{1ex}

     {\centering Standard deviation in parentheses. \\ Significance on 10\% (*), 5\% (**), 1\% (***) level \par}
\end{table}

The result of the regression analysis is largely in line with expectations and achieves a high model quality (resolved variance of about 86\%). The proportion of non-ATCOs, delegated ATM, government ownership, the proportion of domestic flights, traffic volatility, and the technology proxy have a negative impact on ATCO productivity, the latter was not expected, but is statistically insignificant. 

Surprisingly, the complexity indicators have a positive sign, are not significant, or are not relevant. However, this may indicate that the calculation methodology of the complexity values by the PRU does not take into account significant aspects, as discussed in \cite{Standfuss2020c}. Furthermore, the positive correlation between coordination and productivity is counterintuitive, but this interdependency is not statistically significant as well. 

As expected, airspace size has a positive effect on performance. Parabolic and hyperbolic relationships were also tested, but this did not result in higher model quality. In addition, the inclusion of individual FABs was tested. However, this had a negative effect on the results regarding the model quality criteria as well.

The table further shows that despite the variables \textit{DOM} and \textit{AIRP} being correlated (see \autoref{fig:Correl}), both coefficients have the intuitively correct sign and are statistically significant. Further, the VIF-test for collinearity was negative. GDP/capita also shows high correlations to other variables used. In this case, the variable \textit{WEALTH} could not be included in the cross-sectional regression models, also indicated by the VIF-Test. 

Some counter-intuitive results might also be due to the dependent variable: The EUROCONTROL productivity score. Since ANSPs use multiple inputs to "produce" multiple outputs, a purely two-dimensional performance metric might be debatable \cite{Standfuss2022a}. In \cite{Standfuss2021}, DEA was applied to determine efficiency scores for four benchmarking models. We used the scores based on constant returns to scale (CRS). 

In contrast to the PRU score, relative performance, respectively the DEA scores, are limited between 0 and 1. In this case, an OLS regression might not be applicable and regression models with a limited dependent variable are preferable. Whether the results differ significantly is the subject of this investigation. The model quality is maximized for three regression types: OLS, Tobit, and Truncated regression. For this purpose, the DEA scores of models 1, 2, and 2A (published in \cite{Standfuss2021}) are used as dependent variables. \autoref{DEAReg} shows the results after variable reduction. Please note that model 2A consists of 37 ANSPs only, since Maastricht Upper Airspace Control (MUAC) was excluded (see also \cite{Standfuss2021, Standfuss2022}).

\begin{table*}[]
 \caption{Regression Results DEA Score}
 \label{DEAReg}
    \centering
\begin{tabular}{l|ccc|ccc|ccc}
\hline
         & \multicolumn{3}{c}{OLS}                                                                                                                                                                          & \multicolumn{3}{c}{Tobit}                                                                                                                                                                        & \multicolumn{3}{c}{Truncated}                                                                                                                                                                    \\
         & M1                                                             & M2                                                             & M2A                                                            & M1                                                             & M2                                                             & M2A                                                            & M1                                                             & M2                                                             & M2A                                                            \\ \hline 
INT      & \begin{tabular}[c]{@{}c@{}}2,416\\    (0,519)***\end{tabular}  & \begin{tabular}[c]{@{}c@{}}1,716\\    (0,182)***\end{tabular}  & \begin{tabular}[c]{@{}c@{}}-0,961\\    (1,504)\end{tabular}    & \begin{tabular}[c]{@{}c@{}}2,416\\    (0,421)***\end{tabular}  & \begin{tabular}[c]{@{}c@{}}1,716\\    (0,15)***\end{tabular}   & \begin{tabular}[c]{@{}c@{}}-1,181\\    (1,186)\end{tabular}    & \begin{tabular}[c]{@{}c@{}}2,609\\    (1,509)*\end{tabular}    & \begin{tabular}[c]{@{}c@{}}0,74\\    (0,366)**\end{tabular}    & \begin{tabular}[c]{@{}c@{}}0,235\\    (0,972)\end{tabular}     \\ \hdashline
TIME (l) &                                                                &                                                                & \begin{tabular}[c]{@{}c@{}}0,176\\    (0,155)\end{tabular}     &                                                                &                                                                & \begin{tabular}[c]{@{}c@{}}0,192\\    (0,119)\end{tabular}     & \begin{tabular}[c]{@{}c@{}}0,497\\    (0,438)\end{tabular}     &                                                                & \begin{tabular}[c]{@{}c@{}}0,497\\    (0,298)*\end{tabular}    \\
NONA     & \begin{tabular}[c]{@{}c@{}}-0,602\\    (0,199)***\end{tabular} & \begin{tabular}[c]{@{}c@{}}-0,765\\    (0,221)***\end{tabular} & \begin{tabular}[c]{@{}c@{}}-0,333\\    (0,26)\end{tabular}     & \begin{tabular}[c]{@{}c@{}}-0,602\\    (0,162)***\end{tabular} & \begin{tabular}[c]{@{}c@{}}-0,765\\    (0,183)***\end{tabular} & \begin{tabular}[c]{@{}c@{}}-0,39\\    (0,231)*\end{tabular}    & \begin{tabular}[c]{@{}c@{}}-1,184\\    (0,322)***\end{tabular} & \begin{tabular}[c]{@{}c@{}}-0,524\\    (0,184)***\end{tabular} & \begin{tabular}[c]{@{}c@{}}-0,722\\    (0,295)**\end{tabular}  \\
DELATM   &                                                                &                                                                &                                                                &                                                                &                                                                & \begin{tabular}[c]{@{}c@{}}-0,059\\    (0,048)\end{tabular}    &                                                                &                                                                & \begin{tabular}[c]{@{}c@{}}-0,06\\    (0,052)\end{tabular}     \\
MET      & \begin{tabular}[c]{@{}c@{}}0,044\\    (0,041)\end{tabular}     &                                                                &                                                                & \begin{tabular}[c]{@{}c@{}}0,044\\    (0,033)\end{tabular}     &                                                                &                                                                & \begin{tabular}[c]{@{}c@{}}0,099\\    (0,049)**\end{tabular}   &                                                                & \begin{tabular}[c]{@{}c@{}}0,043\\    (0,037)\end{tabular}     \\
AIRP     & \begin{tabular}[c]{@{}c@{}}0,303\\    (0,107)***\end{tabular}  & \begin{tabular}[c]{@{}c@{}}0,18\\    (0,133)\end{tabular}      & \begin{tabular}[c]{@{}c@{}}0,293\\    (0,134)**\end{tabular}   & \begin{tabular}[c]{@{}c@{}}0,303\\    (0,087)***\end{tabular}  & \begin{tabular}[c]{@{}c@{}}0,18\\    (0,11)\end{tabular}       & \begin{tabular}[c]{@{}c@{}}0,287\\    (0,104)***\end{tabular}  & \begin{tabular}[c]{@{}c@{}}0,222\\    (0,117)*\end{tabular}    & \begin{tabular}[c]{@{}c@{}}0,183\\    (0,099)*\end{tabular}    & \begin{tabular}[c]{@{}c@{}}0,312\\    (0,065)***\end{tabular}  \\
JSC      &                                                                &                                                                & \begin{tabular}[c]{@{}c@{}}0,079\\    (0,062)\end{tabular}     &                                                                &                                                                & \begin{tabular}[c]{@{}c@{}}0,084\\    (0,053)\end{tabular}     &                                                                &                                                                &                                                                \\
STATE    & \begin{tabular}[c]{@{}c@{}}-0,078\\    (0,038)**\end{tabular}  & \begin{tabular}[c]{@{}c@{}}-0,116\\    (0,042)**\end{tabular}  & \begin{tabular}[c]{@{}c@{}}-0,073\\    (0,056)\end{tabular}    & \begin{tabular}[c]{@{}c@{}}-0,078\\    (0,031)**\end{tabular}  & \begin{tabular}[c]{@{}c@{}}-0,116\\    (0,035)***\end{tabular} & \begin{tabular}[c]{@{}c@{}}-0,072\\    (0,046)\end{tabular}    & \begin{tabular}[c]{@{}c@{}}-0,128\\    (0,045)***\end{tabular} & \begin{tabular}[c]{@{}c@{}}-0,1\\    (0,035)***\end{tabular}   & \begin{tabular}[c]{@{}c@{}}-0,124\\    (0,037)***\end{tabular} \\ \hdashline
SIZE (l) & \begin{tabular}[c]{@{}c@{}}-0,036\\    (0,028)\end{tabular}    &                                                                & \begin{tabular}[c]{@{}c@{}}0,063\\    (0,036)*\end{tabular}    & \begin{tabular}[c]{@{}c@{}}-0,036\\    (0,023)\end{tabular}    &                                                                & \begin{tabular}[c]{@{}c@{}}0,074\\    (0,03)**\end{tabular}    & \begin{tabular}[c]{@{}c@{}}-0,26\\    (0,08)***\end{tabular}   & \begin{tabular}[c]{@{}c@{}}0,127\\    (0,05)**\end{tabular}    &                                                                \\
OCEAN    &                                                                &                                                                &                                                                &                                                                &                                                                &                                                                &                                                                &                                                                & \begin{tabular}[c]{@{}c@{}}-0,46\\    (0,115)***\end{tabular}  \\
COORD    &                                                                & \begin{tabular}[c]{@{}c@{}}0,021\\    (0,012)*\end{tabular}    & \begin{tabular}[c]{@{}c@{}}0,018\\    (0,014)\end{tabular}     &                                                                & \begin{tabular}[c]{@{}c@{}}0,021\\    (0,01)**\end{tabular}    & \begin{tabular}[c]{@{}c@{}}0,016\\    (0,012)\end{tabular}     &                                                                &                                                                &                                                                \\
L\_AIRP  &                                                                & \begin{tabular}[c]{@{}c@{}}-0,084\\    (0,046)*\end{tabular}   &                                                                &                                                                & \begin{tabular}[c]{@{}c@{}}-0,084\\    (0,038)**\end{tabular}  &                                                                & \begin{tabular}[c]{@{}c@{}}-0,161\\    (0,072)**\end{tabular}  & \begin{tabular}[c]{@{}c@{}}-0,05\\    (0,03)*\end{tabular}     & \begin{tabular}[c]{@{}c@{}}-0,112\\    (0,037)***\end{tabular} \\
OVER     & \begin{tabular}[c]{@{}c@{}}-0,547\\    (0,15)***\end{tabular}  &                                                                & \begin{tabular}[c]{@{}c@{}}0,18\\    (0,164)\end{tabular}      & \begin{tabular}[c]{@{}c@{}}-0,547\\    (0,122)***\end{tabular} &                                                                & \begin{tabular}[c]{@{}c@{}}0,203\\    (0,146)\end{tabular}     & \begin{tabular}[c]{@{}c@{}}-1,051\\    (0,31)***\end{tabular}  &                                                                &                                                                \\
DOM      & \begin{tabular}[c]{@{}c@{}}-1,165\\    (0,398)***\end{tabular} & \begin{tabular}[c]{@{}c@{}}-0,705\\    (0,446)\end{tabular}    & \begin{tabular}[c]{@{}c@{}}-1,16\\    (0,491)**\end{tabular}   & \begin{tabular}[c]{@{}c@{}}-1,165\\    (0,323)***\end{tabular} & \begin{tabular}[c]{@{}c@{}}-0,705\\    (0,369)*\end{tabular}   & \begin{tabular}[c]{@{}c@{}}-1,273\\    (0,388)***\end{tabular} &                                                                & \begin{tabular}[c]{@{}c@{}}-0,817\\    (0,346)**\end{tabular}  &                                                                \\
GINI     & \begin{tabular}[c]{@{}c@{}}-3,034\\    (0,661)***\end{tabular} & \begin{tabular}[c]{@{}c@{}}-2,76\\    (0,628)***\end{tabular}  & \begin{tabular}[c]{@{}c@{}}-2,673\\    (0,793)***\end{tabular} & \begin{tabular}[c]{@{}c@{}}-3,034\\    (0,536)***\end{tabular} & \begin{tabular}[c]{@{}c@{}}-2,76\\    (0,52)***\end{tabular}   & \begin{tabular}[c]{@{}c@{}}-2,733\\    (0,635)***\end{tabular} & \begin{tabular}[c]{@{}c@{}}-4,068\\    (0,927)***\end{tabular} & \begin{tabular}[c]{@{}c@{}}-2,048\\    (0,543)***\end{tabular} & \begin{tabular}[c]{@{}c@{}}-3,928\\    (0,537)***\end{tabular} \\
DENS  &                                                                &                                                                &                                                                &                                                                &                                                                & \begin{tabular}[c]{@{}c@{}}0,009\\    (0,008)\end{tabular}     & \begin{tabular}[c]{@{}c@{}}0,021\\    (0,016)\end{tabular}     &                                                                & \begin{tabular}[c]{@{}c@{}}0,011\\    (0,008)\end{tabular}     \\
VI       & \begin{tabular}[c]{@{}c@{}}0,54\\    (0,392)\end{tabular}      &                                                                &                                                                & \begin{tabular}[c]{@{}c@{}}0,54\\    (0,318)*\end{tabular}     &                                                                & \begin{tabular}[c]{@{}c@{}}0,403\\    (0,375)\end{tabular}     &                                                                &                                                                &                                                                \\
HI       &                                                                & \begin{tabular}[c]{@{}c@{}}-0,394\\    (0,285)\end{tabular}    &                                                                &                                                                & \begin{tabular}[c]{@{}c@{}}-0,394\\    (0,235)*\end{tabular}   &                                                                &                                                                &                                                                &                                                                \\ 
SI       & \begin{tabular}[c]{@{}c@{}}-1,07\\    (0,48)**\end{tabular}    & \begin{tabular}[c]{@{}c@{}}-1,049\\    (0,334)***\end{tabular} & \begin{tabular}[c]{@{}c@{}}-0,736\\    (0,483)\end{tabular}    & \begin{tabular}[c]{@{}c@{}}-1,07\\    (0,389)***\end{tabular}  & \begin{tabular}[c]{@{}c@{}}-1,049\\    (0,276)***\end{tabular} & \begin{tabular}[c]{@{}c@{}}-0,866\\    (0,396)**\end{tabular}  & \begin{tabular}[c]{@{}c@{}}-1,446\\    (0,626)**\end{tabular}  & \begin{tabular}[c]{@{}c@{}}-0,727\\    (0,303)**\end{tabular}  & \begin{tabular}[c]{@{}c@{}}-1,582\\    (0,331)***\end{tabular} \\ \hdashline
COSTS    & \begin{tabular}[c]{@{}c@{}}0,001\\    (0,001)\end{tabular}     & \begin{tabular}[c]{@{}c@{}}0,001\\    (0,001)\end{tabular}     & \begin{tabular}[c]{@{}c@{}}0,002\\    (0,001)***\end{tabular}  & \begin{tabular}[c]{@{}c@{}}0,001\\    (0)*\end{tabular}        & \begin{tabular}[c]{@{}c@{}}0,001\\    (0,001)\end{tabular}     & \begin{tabular}[c]{@{}c@{}}0,001\\    (0,001)**\end{tabular}   & \begin{tabular}[c]{@{}c@{}}0,002\\    (0,001)**\end{tabular}   &                                                                & \begin{tabular}[c]{@{}c@{}}0,002\\    (0,001)***\end{tabular}  \\ 
RES      & \begin{tabular}[c]{@{}c@{}}-0,004\\    (0,001)***\end{tabular} & \begin{tabular}[c]{@{}c@{}}-0,003\\    (0,002)**\end{tabular}  & \begin{tabular}[c]{@{}c@{}}-0,003\\    (0,002)*\end{tabular}   & \begin{tabular}[c]{@{}c@{}}-0,004\\    (0,001)***\end{tabular} & \begin{tabular}[c]{@{}c@{}}-0,003\\    (0,001)***\end{tabular} & \begin{tabular}[c]{@{}c@{}}-0,003\\    (0,001)*\end{tabular}   & \begin{tabular}[c]{@{}c@{}}-0,011\\    (0,003)***\end{tabular} & \begin{tabular}[c]{@{}c@{}}-0,003\\    (0,001)**\end{tabular}  & \begin{tabular}[c]{@{}c@{}}-0,007\\    (0,002)***\end{tabular} \\ \hline 
Adj. R² & 0,71                                                           & 0,6                                                            & 0,65                                                           &                                                                &                                                                &                                                                &                                                                &                                                                &                                                                \\
Akaike   & -52,05                                                         & -46,65                                                         & -44,20                                                         & -50,05                                                         & -44,64                                                         & -38,75                                                         &                                                                &                                                                &                                                                \\
Log-Lik. & 39,02                                                          & 35,32                                                          & 36,10                                                          & 39,02                                                          & 35,32                                                          & 37,38                                                          & 36,06                                                          & 34,75                                                          & 35,55                                                          \\
N        & 38                                                             & 38                                                             & 37                                                             & 38                                                             & 38                                                             & 37                                                             & 38                                                             & 38                                                             & 37                                       \\ \hline
\end{tabular}
\vspace{1ex}

     {\centering Standard deviation in parentheses. Significance on 10\% (*), 5\% (**), 1\% (***) level \par}
\end{table*}

OLS and Tobit regression show nearly identical results when keeping the benchmarking model constant. This means that the coefficients vary between the DEA models, but not between the two regression types. This is due to the fact that no data are censored within the observations (there are neither DEA values larger than 1, nor less than 0). In consequence, coefficients, significance, and standard errors change particularly in DEA model 2A. In contrast, the results of the truncated regression differ significantly. This is, again, particularly visible for DEA model 2A. 

\autoref{DEAReg} further shows that the FAB dummy is not significant in any of the regression and DEA models. "Oceanic" shows significance in the truncated regression for Model 2A only. Conversely, "Horizontal Interactions" as a complexity factor is only relevant in the case of model 2 when applying OLS and Tobit regression, and only significant when applying censored regression. Generally, complexity factors except Speed Interactions are seldom identified as relevant or significant. This finding is in line with the results using productivity as the dependent variable, emphasizing that the complexity indicator is of low suitability. 

Some factors are relevant in each model and also prove significant in some cases. Moreover, these factors are robust since they have the same sign across all models. A higher share of Non-ATCO staff, governmental control of the ANSP (in the sense of non-privatization), an uneven distribution of traffic over time, speed interactions, and the technology proxy affect efficiency negatively. The latter is counter-intuitive, but the low value of the coefficient suggests that the influence is minor. The cost per ATCO hour always has a positive effect, although relevant in eight out of nine cases.

According to the adjusted R$^2$, the highest variance can be resolved for Model 1, followed by Model 2A and Model 2. With reference to the Akaike criterion, however, the goodness of fit of Model 2 is higher than that of Model 2A. OLS regressions generally achieve a better fit than the corresponding Tobit models. These results are confirmed by the log-likelihood value, which is the highest for Model 1 in each case.

The model quality increases when considering \textit{WEALTH}, but the VIF-test shows multicollinearity. When substituting the \textit{NOFAB} variable with individual FABs, the respective variables are significant. However, the model quality decreases and leads to implausible results regarding the other coefficients\footnote{Collinearity issues might be expected}.

\section{Conclusions and way forward}
Efficiency improvements in air traffic are the subject of political and operational discussions. So far, the focus has been primarily on specific aspects such as economies of scale. Due to a rising focus on ANSP inefficiency and sustainability, a holistic approach is required. Our paper contributes  to this goal. It focuses on potential influencing factors, available data, applicable methods, and the empirical evaluation by regression models. 

Depending on the performance metric used, different methods are recommended in the academic literature. If the dependent variable is not limited, OLS regression can be used. Otherwise, Tobit models or truncated regression are recommended. A general decision between Tobit or truncated model depends on the specification (selection of factors to be considered). In most cases, the Tobit model outperforms other regressions. However, particularly due to the high similarity of the results, the application of an OLS regression offers advantages with regard to the applicability, model evaluation, and, if needed, the predictability of the dependent variable. Overall, the results differ in exceptional cases only, so the general effects can be determined with both methods, confirming the results of \cite{Hoff2007}. 

In order to predict the dependent variable as precisely as possible, insignificant factors (determined by the regression) were filtered out to maximize the model quality. Results show that the productivity according to PRU can be estimated more accurately than the DEA scores. This effect is reinforced since the significant variables vary depending on the DEA model. An exception represent \textit{NONA}, \textit{AIRP}, \textit{STATE}, \textit{GINI}, and \textit{SI}, for which the sign and value of the coefficients were similar, independent of the model. Based on these robust results, it can be assumed that these factors have a negative effect on performance. The significance of the variables depends on the economic modeling in the First Stage DEA as well as the units considered\footnote{including or excluding MUAC, as discussed in \cite{Standfuss2022a}}.

The empirical analyses showed that substantial parts of the inefficiency are caused by exogenous, or at least partially exogenous influences. In particular, structural determinants of traffic significantly influence the predictability and thus the required resource buffers. This hampers an efficient allocation of resources, considering ANSPs trade-off between cost reduction and capacity provision. This also emphasizes the necessity of including recommendations to all stakeholders in order to improve the air traffic system.

Due to the small number of observations, the results should be interpreted as a trend. An extension of the data set can be achieved e.g., by the observations of several years, which are evaluated in the form of a panel analysis. Furthermore, essential factors are missing (due to the data), such as the amount and distribution of General Aviation, or the number, location and frequency of military areas. Significant influence has already been demonstrated for forecast quality and short-term traffic fluctuations. A further contribution might be the inclusion of a weather proxy. This might contribute to a higher model quality as well.

This study represents a first step of a holistic approach to influencing factors. In a further study, we will use the benchmarking models proposed by \cite{Standfuss2022a} as a dependent variable for regression analysis. In addition, time effects are a subject of current research. In this context, we will publish an analysis based on panel data. This enables to show efficiency changes over time based on Malmquist Indices, but also to consider unobserved heterogeneity. Economical evaluation of recent global and local events, such as the Corona pandemic or the Russian attack on Ukraine, will be analyzed.

\bibliographystyle{IEEEtran}
\bibliography{references}
\vspace{12pt}
\end{document}